\documentclass[letter]{emulateapj}
\usepackage{natbib}
\usepackage{epsfig}

\usepackage{amssymb}
\usepackage{amsmath}
\usepackage{latexsym}
\usepackage{ifthen}
\usepackage{url}
\usepackage{comment}

\newcommand{\rxte}{\textit{RXTE}}

\newcommand{\degrees}{\,^\circ}

\newcommand{\nudot}{\dot{\nu}}
\newcommand{\nuddot}{\ddot{\nu}}

\newcommand{\psr}[1]
{\ifthenelse{\equal{#1}{0540}}{PSR~B0540$-$69}{\ifthenelse{\equal{#1}{0531}}{PSR~B0531+21}{\ifthenelse{\equal{#1}{crab}}{PSR~B0531+21}{\ifthenelse{\equal{#1}{0537}}{PSR~J0537$-$6910}{\ifthenelse{\equal{#1}{1756}}{PSR~J1756$-$2251}{\ifthenelse{\equal{#1}{1906}}{PSR~J1906+0746}{\ifthenelse{\equal{#1}{1141}}{PSR~J1141$-$6545}{\ifthenelse{\equal{#1}{2127}}{PSR~B2127+11C}{\ifthenelse{\equal{#1}{0751}}{PSR~J0751+1807}{\ifthenelse{\equal{#1}{1713}}{PSR~J1713+0747}{ ???????}}}}}}}}}}}

\newcommand{\change}[1]{#1}

\slugcomment{\today; accepted for publication in the Astrophysical Journal}
\shorttitle{Timing behavior of PSR~B0540$-$69}
\shortauthors{Ferdman et al.}

\begin{document}

\title{Long-term timing and emission behavior of the young Crab-like pulsar PSR~B0540$-$69}

\author{
R.~D.~Ferdman\footnote{rferdman@physics.mcgill.ca}\altaffilmark{1},
R.~F.~Archibald, and
V.~M.~Kaspi
}

\affil{McGill University, Department of Physics, Rutherford Physics Building, 3600 University Street, Montreal, QC, H3A 2T8, Canada}
\altaffiltext{1}{rferdman@physics.mcgill.ca}

\begin{abstract}
We present timing solutions and spin properties of the young pulsar \psr{0540} from analysis of 15.8 years of data from the \textit{Rossi X-Ray Timing Explorer}.  We perform a partially phase-coherent timing analysis in order to mitigate the pronounced effects of timing noise in this pulsar. We also perform fully coherent timing over large subsets of the data set in order to arrive at a more precise solution.  In addition to the previously reported first glitch undergone by this pulsar, we find a second glitch, which occurred at MJD $52927\pm4$, with fractional changes in spin frequency $\Delta\nu/\nu = (1.64 \pm 0.05) \times 10^{-9}$ and spin-down rate $\Delta\nudot/\nudot = (0.930 \pm 0.011) \times 10^{-4}$ (taken from our fully coherent analysis).  We measure a braking index that is consistent over the entire data span, with a mean value $n = 2.129 \pm 0.012$, from our partially coherent timing analysis.  We also investigated the emission behavior of this pulsar, and have found no evidence for significant flux changes, flares, burst-type activity, or pulse profile shape variations.  While there is strong evidence for the much-touted similarity of \psr{0540} to the Crab pulsar, they nevertheless differ in several aspects, including glitch activity, where \psr{0540}
can be said to resemble \change{certain} other very young pulsars.
It seems clear that the specific processes governing the formation, evolution, and interiors of this population of recently born neutron stars can vary significantly, as reflected in their observed properties.
\end{abstract}


\section{Introduction}
Several neutron stars (NSs) are presently observable as pulsars relatively soon after the supernova event associated with their formation.  These objects, which include well studied sources such as the Crab pulsar (PSR~B0531+21) and the Vela pulsar (PSR~B0833$-$45), appear to have been born with relatively high rotation frequencies, and undergo a rapid period of spin-down early in their lives as NSs.  We can generalize the rate at which this rotational energy loss occurs via a power-law model:
\begin{equation}
\dot\nu = K\nu^n,
\label{eqn:rot_power}
\end{equation}
where $\nu$ and $\dot\nu$ are the pulsar spin frequency and frequency derivative, respectively, $K$ is \change{generally} related to the NS moment of inertia, \change{misalignment between the spin and magnetic axes}, and magnetic field strength, and $n$ is the braking index. For the generally utilized case of a \change{pure} magnetic dipole radiation model for pulsar spin-down, $n=3$ \change{for constant $K$} \citep[e.g.,][]{og69}, but alternative models of the pulsar magnetosphere, crust, and/or interior may predict different values for $n$, including a magnetic field that evolves with time \citep[e.g.,][]{br88,mp89}, varying \change{spin-magnetic inclination angle} \citep[e.g.,][]{lgw+13,ljg+15}, or other mechanisms contributing to angular momentum loss \citep{mph01}, such that $K$ in Equation~\ref{eqn:rot_power} is no longer assumed to be constant.

Most pulsars have such a small value of second frequency derivative that its determination is usually not feasible, even over very long timescales.  In contrast, young pulsars are often observed to spin down at a sufficiently fast rate from their birth spin rates that the second derivative of the rotation frequency $\ddot\nu$ is measurable.  In these cases, we may take the derivative of Equation~\ref{eqn:rot_power}, and find that we may cast the braking index as follows:
\begin{equation}
	n = \frac{\ddot{\nu}\nu}{\nudot^2}.
	\label{eqn:braking_index}
\end{equation}
We can therefore determine the braking index $n$ through measurements of the rotation frequency and its derivatives.  There are currently \change{nine} pulsars for which the braking index has been measured; all have been found to have $n<3$ \citep{lps93, lpgc96, mmw+06, lkg+07, lnk+11,  elk+11, wje11, rgl12}.  This indicates that there may indeed be a more complex picture regarding the magnetic field behavior and/or early evolution of the NS.  Continued measurement of pulsar braking indices is therefore a key component to understanding NS magnetospheric behavior.

There exist effects intrinsic to the NS than can influence the reliability of the measurement of $\ddot\nu$, and thus $n$, consequently limiting the number of pulsars for which braking index can be  measured.  We observe \emph{timing noise} as a long-term, quasi-random evolution in pulsar spin frequency \citep[e.g.,][]{hlk10}.  Timing noise is thought to be related to magnetospheric instabilities, having been shown to have a strong relationship to profile changes \citep{lhk+10}.  In contrast, \emph{glitches} are near-instantaneous changes in rotation frequency, and often in spin-down rate, which are believed to result from a sudden coupling of stellar crust and crustal superfluid \citep[e.g.,][and references therein]{elsk11}.  As of this writing, there are 165 pulsars which have been observed to glitch, totaling over 470 recorded glitch events, according to the catalog maintained by Jodrell Bank Observatory pulsar research group\footnote{\url{http://www.jb.man.ac.uk/pulsar/glitches.html}} \citep{elsk11}.

It has been shown that the glitch event rate of pulsars is roughly correlated with age, such that younger pulsars undergo more frequent glitch events that their older counterparts \citep[e.g.,][]{elsk11}. Indeed, very few glitches are seen from pulsars with large characteristic age; for example, only a single glitch has ever been reported from a millisecond pulsar \citep{cb04}.  The very youngest pulsars do, however, appear to show a reduced glitch \change{activity} \citep[see, e.g.,][and references therein]{elsk11}.  One postulated cause for this is the relatively high temperatures in their interiors.  In this scenario, after some cooling time, the pulsars will glitch more often, as is the case with, e.g., the ``adolescent'' Vela pulsar \citep[][]{ml90,lel99,dml02}.

\psr{0540} is among the youngest known pulsars, with a characteristic age $\tau_c = \nu/2\nudot \sim 1700$\,years, \change{and a} rotation frequency $\nu = 19.8$\,Hz, or period $P_\mathrm{spin} = 50$\,ms.  Its X-ray pulsations were discovered by \citet{shh84} in \textit{Einstein Observatory} data associated with the Large Magellanic Cloud supernova remnant 0540$-$693. Its very large spin-down rate not only identified it as a likely young NS, but also made it possible to measure its braking index.  It was soon discovered to have optical pulsations \citep{mp85}, and was immediately identified as very similar to the Crab pulsar, in its photometric and pulsation properties \citep[e.g.,][]{mpb87,hne+02}, the characteristics of its observed radio emission and giant pulse behavior \citep[e.g.,][]{mml+93,jr03,jrmz04}, as well as in the nebula that its wind apparently powers \citep[e.g.,][]{msk93}.

A braking index was first determined for \psr{0540} using both X-ray and optical measurements of the spin frequency and its derivatives, to be $n = 3.6\pm0.8$ \citep{mpb87}.  This measurement disagreed significantly with that of \citet{mp89}, who subsequently found $n=2.01\pm0.02$ using independent observations of optical pulsations from \psr{0540}. Since then, five additional braking index measurements have been made for this pulsar, few of which agree within measurement uncertainties \citep{mpb87,mp89,oh90,dnb99,zmg+01,cmm03,lkg05}. The reason for this is likely due to a combination of timing noise contamination of the output parameter values, and possibly the existence of unaccounted-for glitch activity, particularly in the earlier X-ray and optical data.  Even in studies by \citet{zmg+01} and \citet{cmm03}, both of which performed phase-coherent timing of data from the \textit{Rossi X-ray Timing Explorer} (\textit{RXTE}) to derive pulsar parameters, neither resulted in consistent braking indices, nor did they agree on the cause of observed $\nu$ and $\nudot$ discontinuities near MJD 51325, with the former authors attributing it to a glitch, and the latter to strong timing noise.

The most recent timing analysis of \psr{0540} by \citet{lkg05} re-examined the \rxte{} data sets used by \citet{zmg+01} (1.2 years of \rxte{} observations) and \citet{cmm03} (1.2 years of \rxte{} data; $\sim 8$ years combined data set), and extended them to include 7.6 years of observations.  In addition to a fully phase-coherent analysis, they performed \emph{partially} phase-coherent timing, which divides the data set into shorter, non-overlapping subsets of data in order to determine the change in $\nu$ and $\nudot$ in a way that is more robust against the contamination of parameter determination from timing noise.  They confirmed that there is indeed a glitch at the same epoch claimed by \citet{zmg+01}.  \change{However, they report a significantly different measurement of the braking index of \psr{0540} to that of \citealt{zmg+01}.  The fully coherent analysis performed by \citet{lkg05} on the same data set supports the glitch parameter and braking index measurements determined through their partially coherent analysis, and validating the reliability of that method.}

In this work, we re-analyze the data set used by \citet{lkg05} and extend its span to include all \psr{0540} data taken with \rxte{} until its decommissioning, giving us a 15.8-year baseline for our timing work.  In \S\ref{sec:obs}, we describe the observations and our reduction of the data set, including our time-of-arrival (TOA) calculations.  In \S\ref{sec:timing}, we present out timing efforts, and the results from both partially and fully coherent timing analyses, which we describe in more detail.  We discuss our search for radiative changes, including burst activity in the \rxte{} time series in \S\ref{sec:flux}, and attempt to set limits on emission variations from \psr{0540}.  Finally, we devote \S\ref{sec:discussion} to examining of the implications of our results, including a discussion of the much-published comparisons between \psr{0540} to the Crab pulsar, and how our findings might fit into the overall picture regarding the apparent similarity of these two systems.  We also discuss how such comparisons may contribute to the understanding of the rotational and magnetic field evolution of \psr{0540}, and possibly other NSs like it.

\section{Observations and data reduction}
\label{sec:obs}

All observational data of \psr{0540} used in this work were taken with the Proportional Counter Array \citep[PCA;][]{jmr+06} onboard \rxte{}. The PCA is an array of five proportional counter units (PCUs), which together have a field of view of approximately 1$\degrees$ FWHM and a total collecting area of 6500 cm$^2$.  It is sensitive to an energy range of $2-60$ keV, and records photon arrival times with a  resolution of 1\,$\mu$s.  We used those data acquired in ``GoodXenon'' mode, which simultaneously provides high time resolution with energy information resolved over 256 spectral channels for each event that is not rejected due to background filtering.

Correction of all event data to barycentric dynamical time (TDB) was applied using the \texttt{barycorr} script provided as part of the \rxte{} \textsc{FTOOLS} package\footnote{Provided as part of the HEASOFT software analysis package; version 6.14 was used for this work (see \url{http://heasarc.gsfc.nasa.gov/lheasoft/ftools/fhelp/barycorr.html})}.  Barycentering was performed relative to the pulsar coordinates determined by \citet{msd+10} with observations from the \textit{Hubble Space Telescope}, which remain the most precise position measurement for PSR~B0540$-$69.  For our analysis, we used events found in all PCU layers for our data set with energies within the range $2-18$ keV.

Integrated pulse profiles were constructed for each observing epoch by folding the given time series over 32 phase bins.  This was done by calculating the rotational phase of each event using the rotation frequency and frequency derivative reported by \citet{lkg05} as our \change{reference ephemeris}.  We accumulated a profile in this way for each observing epoch, and assigned a time stamp and pulse phase to the resulting profiles corresponding to the midpoint in time of the given epoch.

To determine pulse times-of-arrival (TOAs) with which to perform our timing analysis, we first constructed a high signal-to-noise ratio (S/N) template profile.  This was done by aligning in rotational phase, then adding together all integrated profiles from our full data set.  TOAs and corresponding uncertainties for a given epoch were then calculated by randomly generating 512 profiles based on the Poissonian count error on each phase bin for the corresponding integrated profile.  We then \change{cross-correlated} each trial profile against the high S/N template in the Fourier domain, in order to calculate a phase shift relative to the fundamental harmonic of the template profile.  The mean shift from all trials for a given epoch was then converted to absolute time offset via the reference pulse period, and added to the time-stamp of each profile to produce a corresponding TOA \citep{tay92}. The corresponding uncertainty for a given TOA was found from the standard deviation of all corresponding trial TOAs used for that observing epoch.
\change{Profile smearing effects on later-epoch TOA uncertainties from folding event data with our reference ephemeris are negligible; the fractional rotation frequency error $\delta\nu/\nu$ is orders of magnitude smaller than the precision required to avoid significant spin phase drift within a profile bin, over the typical integration times of $\lesssim 5000$\,s.}
Uncertainties therefore reflect the differences in profile S/N, which is principally due to the different integration times, \change{as well as the number of active PCUs}, throughout the data set.

\section{Timing Analysis}
\label{sec:timing}

In this work, we have extended the \citet{lkg05} data set by more than a factor of two, to include 15.8 years of \rxte{} data.  We have performed a phase-coherent timing analysis on these data, which allows us to account for every rotation of the NS.  PSR~B0540$-$69 is an isolated pulsar; phase coherence can be achieved by fitting the measured pulse TOAs to a spin model that determines the rotational phase $\phi$ of the pulsar at a given time $t$, via a Taylor expansion of the spin frequency of the pulsar and its derivatives:
\begin{equation}
	\phi(t) = \phi(t_0) + \nu_0(t - t_0) + \frac{1}{2}\nudot_0(t - t_0)^2 + \frac{1}{6}\nuddot_0(t - t_0)^3 + \ldots
\end{equation}
Here, $t_0$ is a reference epoch in our model, and $\nu_0$, $\nudot_0$, and $\nuddot_0$ represent the pulsar spin frequency and its derivatives at epoch $t_0$.

For all timing analyses, we use the \textsc{tempo2} software package \citep{ehm06, hem06}, which fits our barycentered TOAs to a model describing the spin evolution of the pulsar.
As with our pulse profile construction, we used the spin frequency and frequency derivatives from the \citet{lkg05} model ephemeris as a starting point for our timing analyses. We have modified the pulsar position to match that of \citet{msd+10} mentioned above, and have updated the Solar System model to use the JPL DE421 ephemeris \citep{sta04b} in order to account for the motion of the Earth.  We have not included any parameters describing \textit{a priori} knowledge of glitch behavior of this pulsar, allowing for an independent analysis and comparisons with previous results.

Following the general procedure used by \citet{lkg05}, we performed timing analysis in two independent ways: in \S{\ref{sec:part_coherent}}, we describe our effort to reduce the effects of intrinsic pulsar spin noise by performing a partially coherent analysis, in which we divide the data set into relatively short timing intervals; in \S{\ref{sec:coherent}}, we describe our fully coherent timing analysis, in which we attempt to phase-connect the entire data set (over two large overlapping sections), and compare our results from both methods.

\begin{figure*}
  \begin{center}
    \includegraphics[width=1.0\textwidth]{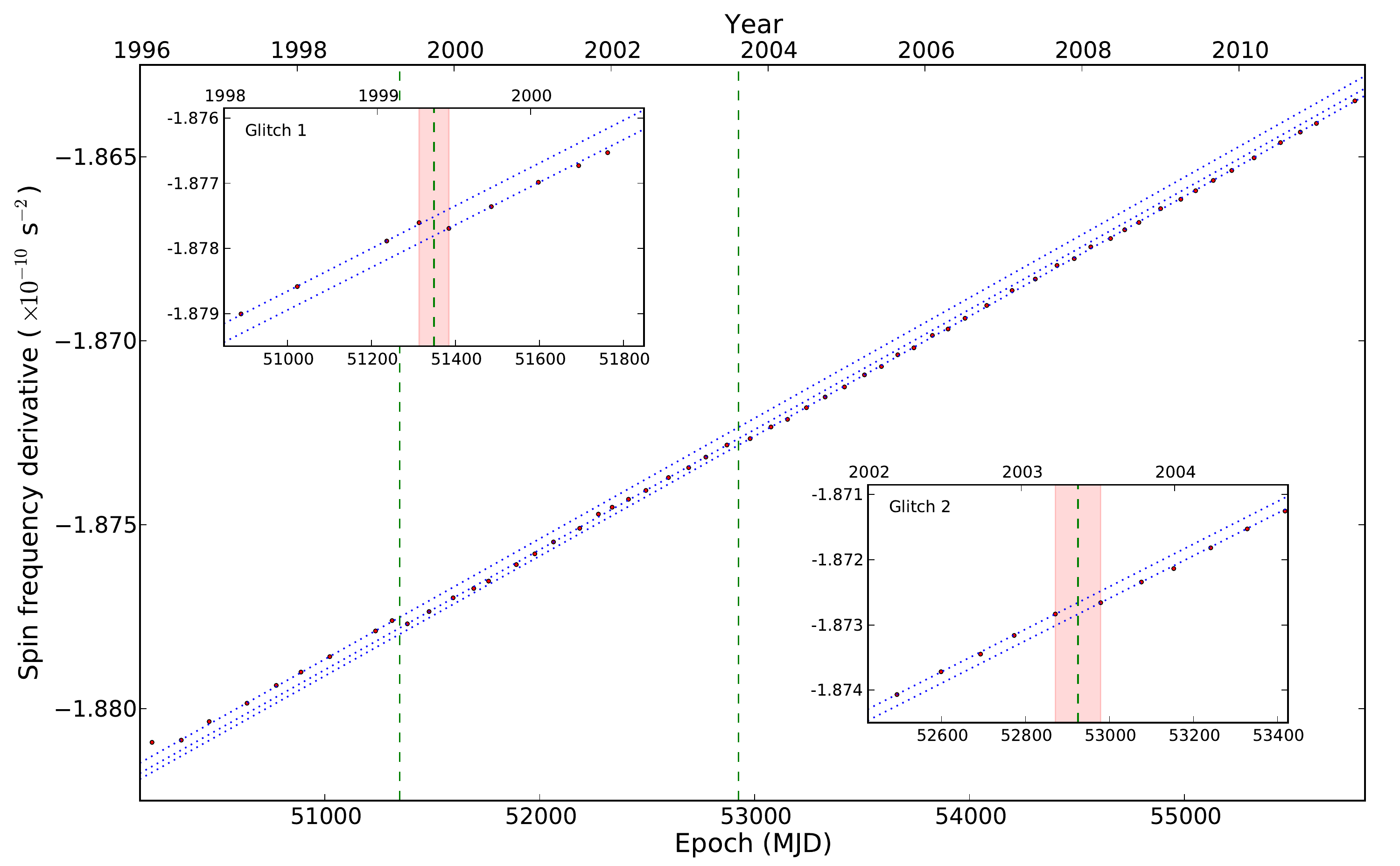}
    \caption{Partially coherent timing measurements of the spin frequency derivative $\nudot$ of PSR~B0540$-$69. Points show fit values of $\nudot$ for each data subset.  Vertical dashed green lines show the glitch epochs at MJDs $51349 \pm 36$ and $52925 \pm 54$.  Blue dashed lines show fit slopes to $\nudot$ data before the first glitch, between glitches, and after the second glitch.  Insets focus on a portion of the data near each glitch, with red shaded areas showing the extent of the possible glitch epoch.\label{fig:part_coherent_nudot}}
  \end{center}
\end{figure*}

\subsection{Partially Phase-coherent Analysis}
\label{sec:part_coherent}

Young pulsars often exhibit high levels of timing noise that can inhibit accurate measurement of the intrinsic frequency derivatives, and thus braking indices, as well as identification and characterization of glitches.  While phase-coherent timing can provide very precise parameter determinations, higher-order spin derivatives are often needed to remove the effects of spin noise.  This is typically a red process, occurring over long timescales, leading to covariances between these parameters.

In order to mitigate the influence of timing noise, we divided the full data set into 59 non-overlapping subsets.  The length of each subset was determined based on its timing properties; specifically, we included the number of TOAs that allowed for a phase-coherent timing fit that required only the rotational frequency $\nu$ and its first derivative $\nudot$ to maintain white, Gaussian-distributed post-fit timing residuals.  For each subset, we have taken the center of the time span as the reference epoch for the timing fit.

We find evidence for a glitch at MJD $51348 \pm 36$, consistent with that found by \citet{lkg05} and \citet{zmg+01}.  In this analysis, we quote the midpoint between the epochs just before and after the \change{discontinuity found in $\nudot$} as the value of the glitch epoch.  The (symmetric) uncertainty is simply calculated to be the distance in time between the glitch and either neighboring epochs.  In our extended data set, we have also found a second glitch at MJD $52925 \pm 54$.  In Figure~\ref{fig:part_coherent_nudot} we plot our measurements of the spin frequency derivative at each subset epoch for this partially coherent analysis; both glitches are denoted by a vertical dotted line at their respective epochs, and are the focus of the inset plots.  Finally, we find some evidence for a previously unreported glitch near the start of the data set, which cannot be corroborated due to a lack of earlier data points.  If this is a true glitch, we determine that it would have occurred at MJD $50264 \pm 68$.

We were able to fit three distinct slopes to our measurements of $\nudot$, allowing us to determine $\nuddot$ for three separate date ranges: before the first glitch (not including the first epoch); between our two confirmed glitch events; and after the second glitch.  We determined uncertainties for our $\nuddot$ measurements through a bootstrap technique.  Here, we randomized the data order many times to construct a distribution of fit $\nuddot$ values, the width of which was taken to be the uncertainty in $\nuddot$.  This analysis showed the errors taken directly from the output covariance matrices from our linear fits to be underestimated, and therefore report our bootstrap-derived uncertainties in Table~\ref{tab:0540_params}.
\change{Our measurements of $\nuddot$ are consistent within their mutual $1\sigma$ uncertainties; we therefore claim no significant third frequency derivative measurement can be made from our analysis of this data set.}

\begin{deluxetable*}{lcccccc}
    \tablecolumns{7}
    \tablecaption{Timing parameters for PSR~B0540$-$69\label{tab:0540_params}}
    \tablewidth{0pc}
\startdata 
\hline\hline
\multicolumn{7}{c}{Data set} \\
\hline
Data span (yr)\dotfill & \multicolumn{6}{c}{15.8} \\
Date range (MJD)\dotfill & \multicolumn{6}{c}{$50123.2 - 55898.7$} \\ 
Number of TOAs\dotfill & \multicolumn{6}{c}{2247} \\
Solar system ephemeris model\dotfill & \multicolumn{6}{c}{DE421} \\ 
\hline
\multicolumn{7}{c}{Set Quantities} \\ 
\hline

R.~A.\tablenotemark{a} (J2000), $\alpha$\dotfill & \multicolumn{6}{c}{05:40:11.202} \\ 
Decl.\tablenotemark{a} (J2000), $\delta$\dotfill & \multicolumn{6}{c}{$-$69:19:54.17} \\ 
\hline
\hline
\multicolumn{7}{c}{Measured Quantities---Partially Coherent Timing Fit}\\ 
\hline
    &   \multicolumn{2}{c}{Subset 1} & \multicolumn{2}{c}{Subset 2} & \multicolumn{2}{c}{Subset 3} \\ 
\hline
Date range (MJD) & \multicolumn{2}{c}{$50296.9-51333.1$} & \multicolumn{2}{c}{$51342.9-52921.0$} & \multicolumn{2}{c}{$52930.6-55830.9$} \\
Second derivative of rotation frequency, $\ddot{\nu}$ ($\times 10^{-21}$ s$^{-3}$)\dotfill & \multicolumn{2}{c}{3.79(3)}  &  \multicolumn{2}{c}{3.78(2)}  &  \multicolumn{2}{c}{3.773(6)}\\
Braking index, $n$\dotfill & \multicolumn{2}{c}{2.127(18)}  &  \multicolumn{2}{c}{2.127(13)}  &\multicolumn{2}{c}{2.135(4)} \\
\\
Glitch epoch (MJD)\dotfill                            & \multicolumn{3}{c}{\hspace{1.2cm} 51349(36)}   &   \multicolumn{3}{c}{\hspace{-1.3cm} 52925(54)}  \\
$\Delta{\dot\nu}\ (\times 10^{-14}$ s$^{-2})$\dotfill & \multicolumn{3}{c}{\hspace{1.2cm} -2.9(2)}     &  \multicolumn{3}{c}{\hspace{-1.3cm} -1.77(15)}\\
$\Delta{\dot\nu}/\dot\nu\ (\times 10^{-4})$\dotfill   & \multicolumn{3}{c}{\hspace{1.2cm} 1.554(6)}    & \multicolumn{3}{c}{\hspace{-1.3cm} 0.9460(11)}\\
\hline
\hline
\multicolumn{7}{c}{Measured Quantities---Coherent Timing Fit} \\ 
\hline
    &   \multicolumn{3}{c}{\hspace{1.2cm} Subset 1} & \multicolumn{3}{c}{\hspace{-1.3cm} Subset 2} \\ 
\hline
Date range (MJD) & \multicolumn{3}{c}{\hspace{1.2cm} $50123.2 - 52935.7$} & \multicolumn{3}{c}{\hspace{-1.3cm} $51342.9 - 55898.7$} \\ 
Rotation frequency, $\nu$ (s$^{-1}$)\dotfill & \multicolumn{3}{c}{\hspace{1.2cm} 19.8024438758(9)} &  \multicolumn{3}{c}{\hspace{-1.3cm} 19.7746860321(7)}\\ 
First derivative of rotation frequency, $\dot{\nu}$ ($\times 10^{-10}$ s$^{-2}$)\dotfill & \multicolumn{3}{c}{\hspace{1.2cm} $-$1.8780249(6)}   &  \multicolumn{3}{c}{\hspace{-1.3cm} $-$1.8727175(8)}  \\ 
Second derivative of rotation frequency, $\ddot{\nu}$ ($\times 10^{-21}$ s$^{-3}$)\dotfill & \multicolumn{3}{c}{\hspace{1.2cm} 3.721(10)}   &  \multicolumn{3}{c}{\hspace{-1.3cm} 3.772(5)}\\ 
Reference timing epoch (MJD)\dotfill & \multicolumn{3}{c}{\hspace{1.2cm} 51197}   &  \multicolumn{3}{c}{\hspace{-1.3cm} 52910}\\
Braking index, $n$\dotfill & \multicolumn{3}{c}{\hspace{1.2cm} 2.08(3)}   &  \multicolumn{3}{c}{\hspace{-1.3cm} 2.12(3)}\\
RMS timing residual (ms)\dotfill & \multicolumn{3}{c}{\hspace{1.2cm} 0.74}   &  \multicolumn{3}{c}{\hspace{-1.3cm} 1.75}\\
$\chi^2$ of fit\dotfill & \multicolumn{3}{c}{\hspace{1.2cm} 1293.22} & \multicolumn{3}{c}{\hspace{-1.3cm} 10088.45} \\
Degrees of freedom\dotfill & \multicolumn{3}{c}{\hspace{1.2cm} 719}  & \multicolumn{3}{c}{\hspace{-1.3cm} 1432} \\
Number of fit frequency derivatives\dotfill & \multicolumn{3}{c}{\hspace{1.2cm} 12}    &  \multicolumn{3}{c}{\hspace{-1.3cm} 12}\\
\\
Glitch epoch (MJD)\dotfill & \multicolumn{3}{c}{\hspace{1.2cm} 51335(13)}   &  \multicolumn{3}{c}{\hspace{-1.3cm} 52927(4)}\\
$\Delta{\nu}\ (\times 10^{-8}$ s$^{-1})$\dotfill & \multicolumn{3}{c}{\hspace{1.2cm} 2.5(6)}   &  \multicolumn{3}{c}{\hspace{-1.3cm} 3.24(10)}\\
$\Delta{\nu}/\nu\ (\times 10^{-9})$\dotfill & \multicolumn{3}{c}{\hspace{1.2cm} 1.3(3)}   &  \multicolumn{3}{c}{\hspace{-1.3cm} 1.64(5)}\\
$\Delta{\dot\nu}\ (\times 10^{-14}$ s$^{-2})$\dotfill & \multicolumn{3}{c}{\hspace{1.2cm} $-$2.75(7)}   &  \multicolumn{3}{c}{\hspace{-1.3cm} $-$1.74(2)}\\
$\Delta{\dot\nu}/\dot\nu\ (\times 10^{-4})$\dotfill & \multicolumn{3}{c}{\hspace{1.2cm} 1.46(4)}   &  \multicolumn{3}{c}{\hspace{-1.3cm} 0.930(11)}
\enddata
\tablenotetext{a}{Adopted from \citet{msd+10}}

\end{deluxetable*}

\begin{figure}
  \begin{center}
    \includegraphics[width=0.5\textwidth]{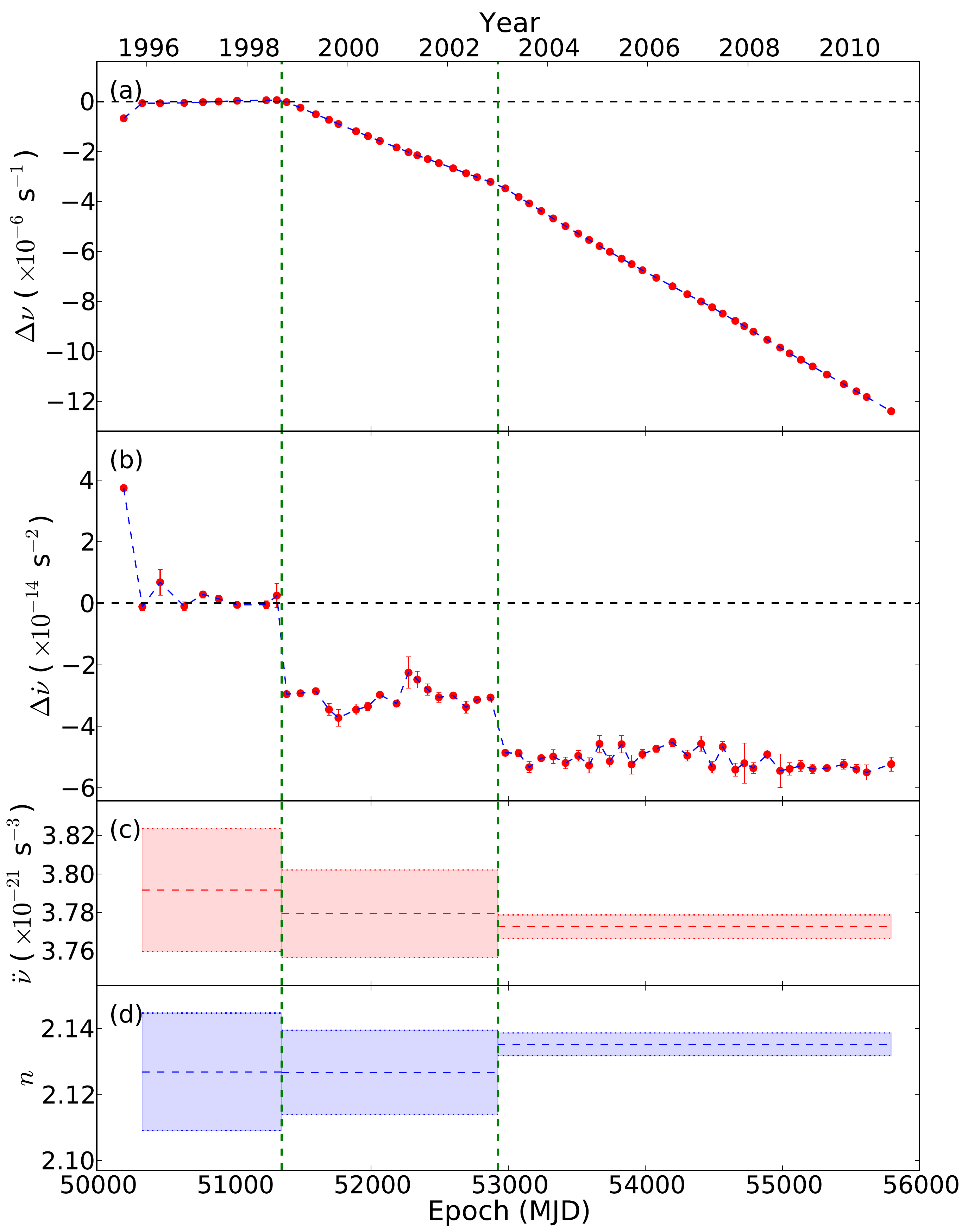}
    \caption{Summary of partially coherent timing analysis of PSR~B0540$-$69. Panel (a): residual measurements of rotation frequency, \change{determined by removing our reference model, which is valid prior to the first reported glitch (fiducial epoch MJD 51197)}.  Discontinuities are seen to occur around the time of each glitch.  Panel (b): residual spin-down rate \change{where the fit slope prior to the first glitch is first removed.}  Clear offsets are  seen to be associated with the reported glitches, as well as persistent spin-down rate increases after each glitch.  Panels (c) and (d): second rotation frequency derivative and braking index, respectively. The values (dashed line) and uncertainties (dotted lines) shown were determined using data taken before the first glitch, between the two reported glitches, and after the second glitch; these date ranges represented as colored regions.\label{fig:part_coherent_summary}}
  \end{center}
\end{figure}

With our measured $\nu$, $\nudot$, and $\nuddot$, we were then able to calculate a mean braking index for each epoch range,
finding $n = 2.127 \pm 0.017$, $2.127 \pm 0.013$, $2.135 \pm 0.004$ for the first, second and third data spans, respectively.  These are consistent over the data set within $1\sigma$; we therefore quote a mean value $n = 2.129 \pm 0.012$, also consistent with the reported value by \citet{lkg05}.
\change{For each glitch, we calculate the change in frequency derivative $\Delta\nudot$ by determining the difference in the extrapolated $\nudot$ from the partially coherent solutions before and after the glitch epoch.}
From these, we find a fractional change $\Delta\nudot/\nudot = (1.554 \pm 0.006) \times 10^{-4}$ at the first glitch epoch, which is consistent with the value reported by \citet{lkg05}.
At the second glitch epoch, we find $\Delta\nudot/\nudot = (9.460 \pm 0.011) \times 10^{-5}$.
In Figure~\ref{fig:part_coherent_summary}, we plot a summary of our partially coherent timing analysis.  The top two panels show the residual values for $\nu$ and $\nudot$, after their \change{respective trends prior to the first glitch} have been subtracted out.  In the case of $\nu$, we note that one can easily see the measurement discontinuities near the reported glitch epochs (including the proposed, but unconfirmed, initial glitch).
We cannot, however, make a significant measurement of the intrinsic changes in rotation frequency $\Delta\nu$ due to their small magnitudes.
\change{This is because the overall measured frequency change for both glitches is dominated by the change in frequency derivative, over and above the frequency step itself, on a timescale that is significantly shorter than the time between the epochs surrounding each glitch.}
The bottom two panels of Figure~\ref{fig:part_coherent_summary} show the measured values of $\nuddot$ and corresponding braking indices for each of the three epochs, which are delimited by the two reported glitches.

\subsection{Fully Phase-coherent Analysis}
\label{sec:coherent}

\begin{figure*}
  \begin{center}
    \includegraphics[width=\textwidth]{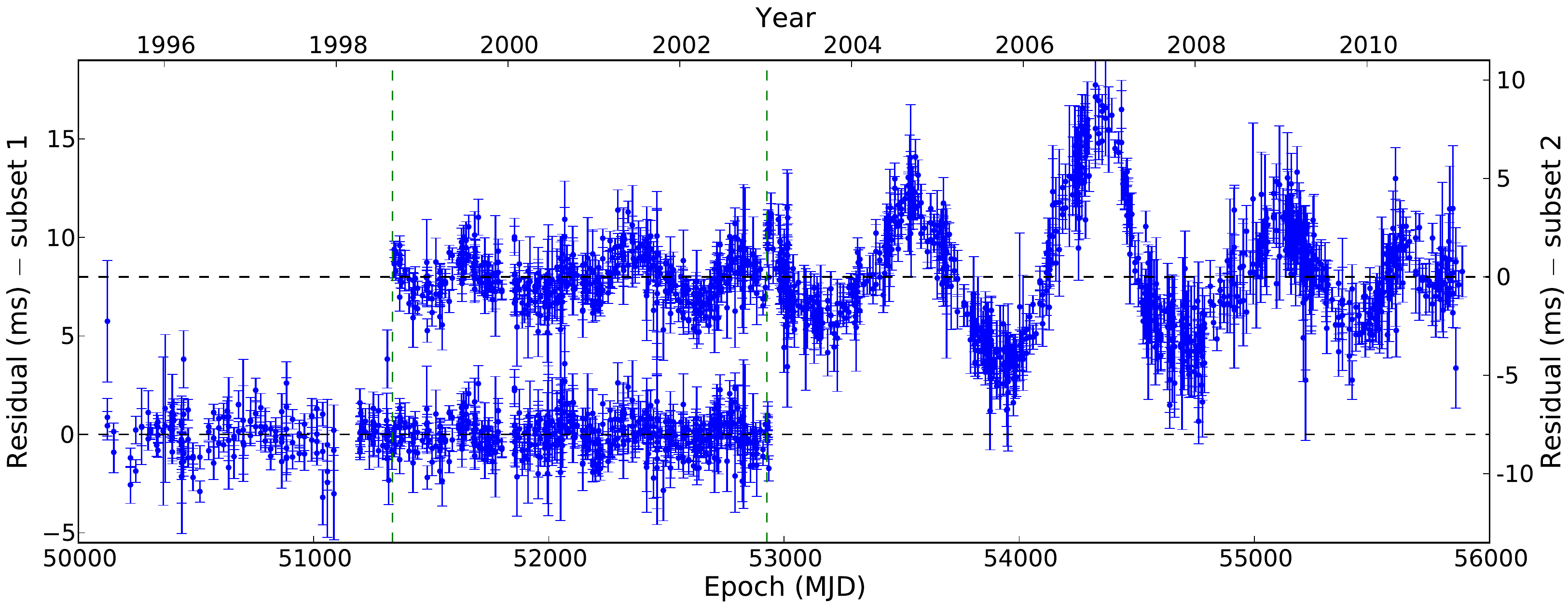}
    \caption{Fully phase-coherent timing residuals for PSR~B0540$-$69, divided into two overlapping subsets.  The first of these includes all data before the second reported glitch, and the second subset comprises all data after the first glitch.  Scales for the first and second timing analysis subsets are shown on the left and right vertical axis labels, respectively.  Glitch epochs are denoted by vertical dashed lines.\label{fig:fully_coherent_resids}}
  \end{center}
\end{figure*}

We performed a coherent timing analysis of our entire data set, in order to confirm the results of our partially coherent analysis.  More importantly, it allowed us to obtain more precise measurements of the glitch epoch, and to \change{determine the change in the pulsar rotation frequency, $\Delta\nu$ (and therefore derive the fractional change $\Delta\nu/\nu$) at each glitch epoch}, which we were not able to achieve in our partially coherent analysis.

As stated in the previous subsection, long-term timing of young pulsars is often affected by a great deal of red spin noise, and this is certainly the case for \psr{0540}. In fact, it was not possible to phase-connect the entire data set with an RMS of the timing residuals within a pulse period. This was due to the need for more than 12 frequency derivatives to do so, which would exceed machine representation capabilities of those values, and is thus not available in \textsc{Tempo2}.  We thus divided the data set into two overlapping segments: the first subset includes all data up to, but not including, the second glitch, and the second subset includes all data after, but not including, the first reported glitch.

Figure~\ref{fig:fully_coherent_resids} shows the post-fit timing residuals for both subsets of the full data set, with an artificial vertical offset added between the two sets for clarity.  In both cases we have included several high-order frequency derivatives in our fit in an attempt to ``whiten out'' as much as possible the effects of red timing noise; specifically, we employed 12 derivatives for both subsets.  A summary of our coherent analysis results can be found in Table~\ref{tab:0540_params}.
\change{We report second frequency derivatives from each subset that are inconsistent by an amount greater than their mutual $3\sigma$ uncertainties; however, our use of high-order frequency derivatives due to timing noise effects has likely introduced covariances between spin parameters that would affect our $\nuddot$ measurements.  This also casts doubt on coherent-based determination of a third derivative $\dddot\nu$ that clearly represents its long-term spin evolution, rather than red-noise effects.}

Our analysis resulted in a braking index measurement for each subset, evaluated at each corresponding reference epoch, also reported in Table~\ref{tab:0540_params}.  Simply adopting the measurement errors based on the formal fit uncertainties would likely underestimate the resulting propagated uncertainty on the braking index $n$; this is due to timing noise contamination that results in covariances between higher-order frequency derivatives. We therefore follow the more conservative method taken by \citet{lkg05}, by taking the uncertainties on each $n$ to be the standard deviation of the calculated $n$ resulting from a set of timing fits that use between 2 and 12 frequency derivatives.  This better reflects the sensitivity of $n$ to the number of parameters in our fit.  We find $n = 2.08 \pm 0.03$ and $2.12\pm0.03$ for the first and second subsets, respectively, in agreement with each other, and with the results of our partially coherent timing within $1\sigma$ uncertainties.  This consistency between subsets and analyses gives us confidence in the unchanging nature of the braking index over the nearly 16 years span of this data set.  We therefore quote a mean braking index $n = 2.10 \pm 0.03$ from this fully coherent timing analysis.

We were additionally able to confirm both glitch epochs found with our partially coherent analysis, and measure the fractional changes in both rotation frequency and spin-down rate through coherent timing analysis.  As expected, these glitches have resulted in low-level changes in $\nu$; we find $\Delta\nu/\nu = 1.3\pm0.3 \times 10^{-9}$ and $1.64\pm0.05 \times 10^{-9}$ for the first and second glitches, respectively; we also find $\Delta\nudot/\nudot = 1.46\pm0.04 \times 10^{-4}$ and $0.930\pm0.011 \times 10^{-4}$.  While our determined fractional change in the rotation frequency agrees well within $1\sigma$ with the reported value of \citet{lkg05}, this is not quite the case for the measured fractional spin-down increase, \change{only agreeing near to their mutual $2\sigma$ ranges;
however, our spin-down glitch sizes $\Delta\nudot$ do agree with those determined from our own partially coherent solutions.}
This is not entirely surprising, given the high level of timing noise seen in the post-fit residuals from the coherent timing analysis (Figure~\ref{fig:fully_coherent_resids}), which results in glitch parameter values that vary with the number of spin frequency derivatives included in our timing model.  We did not find a glitch at the beginning of this data set, near MJD 50264; this may be due to both a lack of data around that epoch and the effects of timing noise, or else because the glitch did not in fact occur.

\section{Searching for Radiative Changes}
\label{sec:rad_change}

We conducted an analysis of our data set to determine whether there existed any flux variations, and if so, whether these could be associated with the observed glitches.
To do so, we first filtered our event data in order to exclude known high-background epochs that may have otherwise biased our flux measurements.  This included the rejection of data during South Atlantic Anomaly passage, Earth occultation and bright Earth effects, and/or electron contamination.  We also rejected epochs with pointing offsets greater than 0.02$\degrees$.  Finally, we did not use any data taken with PCU0 on or after 2000 May 12 (MJD 51676), or with PCU1 on or after 2006 December 25 (MJD 54094) due to loss of their respective propane layers \citep[see, e.g.,][as well as the RXTE guest observer online facility\footnote{\url{http://heasarc.gsfc.nasa.gov/docs/xte/}}]{jmr+06, chp+10}; while this does not affect our TOA calculations and timing analysis, it does influence the calibrated flux values, causing the interpretation of PCU0 and PCU1 flux data to be unreliable after the above dates.

\subsection{Flux Measurements}
\label{sec:flux}

\begin{figure}
  \begin{center}
    \includegraphics[width=0.5\textwidth]{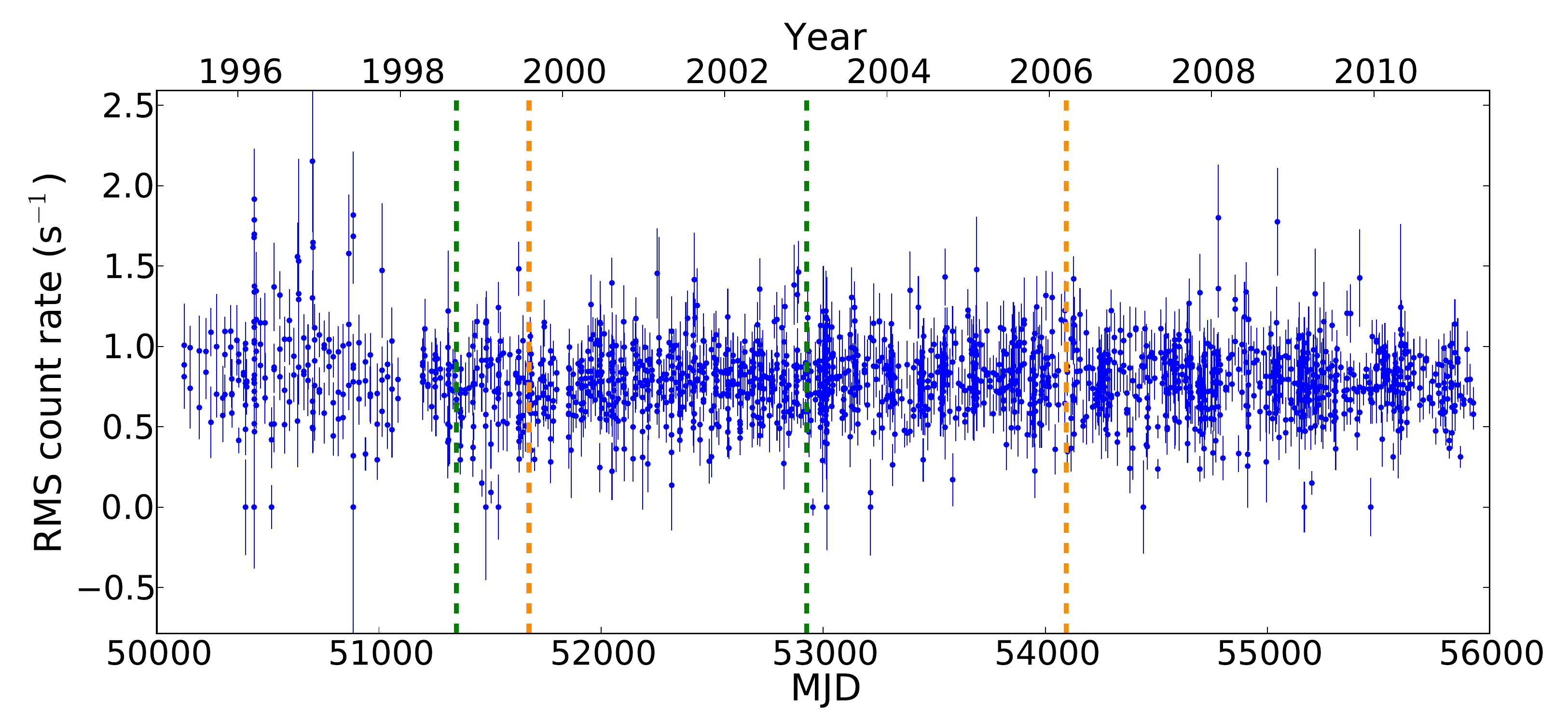}
    \caption{RMS pulsed fluxes for observations of PSR~B0540$-$69, for the energy range $2-18\,$keV; shown are flux data from PCU2.  Green dashed lines show glitch epochs, and orange dotted lines show the epochs at which PCU0 (MJD 51676) and PCU1 (MJD 54094) lost their respective propane layers.\label{fig:fluxes}}
  \end{center}
\end{figure}

We calibrated our flux data separately for each PCU based on the collimator response measured for the Crab pulsar as a function of angular separation between the pulsar position and the nominal telescope pointing \citep[as measured by][]{jmr+06} at each given epoch.  In calculating fluxes, we determined the pulsed count rate for each PCU using the RMS flux variations.  This method \citep[see][for a more detailed description]{ahk+13,abp+14} results in flux measurements with lower bias than the usual method of integrating the pulse profile above the background levels; the latter relies on the uncertainty in the perceived minimum of the profile, which results in larger biases for noisier profiles. In Figure~\ref{fig:fluxes} we show the measured RMS pulsed flux over the data set for PCU2.  We find that there are no significant flux enhancements that are clearly associated with the glitch events reported in \S\ref{sec:timing}.

\subsection{Profile shape changes}
In order to test for changes in profile morphology, we used the template profile we created for our timing analysis described in \S\ref{sec:timing} as a model for the profile shape of \psr{0540}. We first accumulated pulse profiles from data taken within each of the subsets used for the partially coherent analysis described in \S\ref{sec:part_coherent}.  This was done in order to obtain high S/N profiles, while allowing us to have a reasonably sized sample of resulting pulse profiles over the time span of this data set.
For each subset profile, we scaled it to match the template after removing the mean value of both profiles, in order to account for any relative flux changes.  We then aligned each profile in pulse phase with the template, through a frequency-domain cross-correlation similar to that outlined in \S\ref{sec:obs} for TOA calculation.

We determined the goodness-of-fit of our template profile to each aligned and scaled added profile by calculating the $\chi^2$ statistic.  In each case, we find values that are consistent with a $\chi^2$ distribution with the relevant number of degrees of freedom; this includes profiles representing epochs just before and after glitches, which have fits that yield $p$-values greater than 0.99.  We therefore detect no significant change in profile shape over our entire data site, including at those epochs related to the observed glitches.  Based on the lack of measurable flux or profile shape changes, and using the measured glitch epochs determined from coherent timing, we find the maximum time for radiative changes, and subsequent decay, resulting from a glitch to be 21 and 13 days for the first and second glitches, respectively.

\subsection{Search for Burst Activity}
\label{sec:bursts}

We also conducted a search for bursts over the nearly 16\,years for which we have \psr{0540} data.  In order to do so, we created a time series from the event data using $1/32$-s time bins, in the same $2-18\,$keV energy range with which our timing analysis was conducted.  Assuming a Poisson distribution for our time-series data, we looked for individual data points that were outliers from running means over 10, 100, and 1000\,s subsets of our time series, using incremental steps of $50\%$ of each of these timescales \citep[see][where descriptions of our employed method are given]{gkw04, sk11}.  We did not find any significant burst activity in this data set.

The lack of X-ray flux variation, burst activity, or profile changes in \psr{0540} is consistent with the overall observed behavior of most rotation-powered pulsars, which have stable fluxes and pulse shapes.  Exceptions to this are
PSR~J1846$-$0258 \citep{ggg+08,kh09,lkg10,akb+15} and PSR~J1119$-$6127 \citep{wje11,awe+15}, \change{which have an approximately order-of-magnitude larger dipolar magnetic fields than \psr{0540})},
and have displayed radiative changes near a glitch. This suggests a possible link to magnetars, which also often show flux variations connected to observed glitches \citep{dkg07,dkg08,dk14}.  The young, high-field pulsar PSR~B1509$-$58 has shown no glitch activity in over 28 years \citep{lk11}, however, if one is revealed in future observations it could help to shed important light on the properties of young pulsars with high magnetic fields relative to other observed young NSs.

\section{Discussion}
\label{sec:discussion}

\subsection{Braking Index}

Our partially and fully coherent analyses of \psr{0540} result in consistent determination of braking indices, which appear to remain constant throughout our 15.8-yr data set, within measured uncertainties (see Table~\ref{tab:0540_params}).  Furthermore, our measured values of braking index are consistent within $1\sigma$ of the quoted values of $n=2.149\pm0.009$ reported by \citet{lkg05}, and $n=2.125\pm0.001$ by \citet{cmm03}; our values are not consistent  at the mutual $3\sigma$ level with that reported by \citet{zmg+01}, which was likely contaminated by the effects of timing noise.

As with all published braking index values, that of \psr{0540} is significantly lower than the value of 3 predicted for an ideal magnetic dipole field, indicating a departure from the standard model of dipole radiation for pulsar slow-down.   Unlike studies of the Crab pulsar and PSR~B1509$-$58, for example \citep{lkgm05,ljg+15}, we have unfortunately not been able to measure a reliable third frequency derivative $\dddot\nu$ for \psr{0540} \change{(see \S\ref{sec:part_coherent} and \S\ref{sec:coherent})}, which could have given further insight into the possible time variation of its magnetic field.  However, recent modeling of young pulsar braking indices shows that the known braking indices may be explained by higher-order field evolution due to Hall drift in the NS crust, causing an enhancement of the dipole field component \citep{gc15}.  Further discoveries of pulsars with measurable braking indices are therefore crucial for defining a better picture of the magnetic field evolution of young pulsars.

\subsection{Glitches or timing noise?}

\change{There are several factors that give us confidence that our reported events are indeed glitches: (a) our confirmation of the glitch found by \citet{lkg05} near MJD 51348, as well as the consistency in its corresponding measured fractional change in $\nu$, lends credence to its existence, rather than attributing it to timing noise effects, as was postulated by \citet{cmm03}; (b) our discovery of a second glitch near MJD 52925 of similar fractional change in both $\nu$ and $\nudot$ to that of the first glitch demonstrates consistent behavior; and (c) the persistent offsets observed in spin-down rate after each glitch leads us to favor the interpretation of these discontinuities as genuine glitches.}

We can compare this behavior to that of the Crab and other young pulsars.  Although some have experienced higher glitch event rates, some have undergone many low-level glitches, several of which have smaller fractional change in both $\nu$ and $\nudot$ than those seen in \psr{0540}.  Figure~\ref{fig:glitch_compare} plots $\Delta\nu/\nu$ and $\Delta\nudot/\nudot$ for \psr{0540}, as well as those compiled by \citet{elsk11} for the Crab and other pulsars. The glitch sizes we find in the \psr{0540} data do not seem to occupy a particularly unique part of parameter space, and are well within those glitch event sizes found for many young pulsars.  \change{Finally, as discussed in the following section, the Crab pulsar is also known to show persistent steps in frequency derivative.}

\subsection{A Crab twin?}
\label{sec:crab_twin}

Since the discovery of \psr{0540}, much has been discussed of its resemblance to the Crab pulsar \citep[][]{shh84}, and their similarity in several properties has supported this comparison. For the Crab, its characteristic age $\tau_{c,\mathrm{Crab}} = 1258$\,years, measured spin luminosity $\dot{E}_\mathrm{Crab} = 4.5\times10^{38}$\,erg\,s$^{-1}$, and surface magnetic field strength $B_\mathrm{s,Crab} = 3.8\times10^{12}$\,G, are indeed tantalizingly close to those of \psr{0540}, which has $\tau_{c,\mathrm{0540}} = 1672$\,years, $\dot{E}_\mathrm{0540} = 1.5\times10^{38}$\,erg\,s$^{-1}$, and $B_\mathrm{s,0540} = 5.0\times10^{12}$\,G, hinting at a similar formation and evolution.  Additionally, as discussed above, the Crab has experienced many low-level glitches events in both $\nu$ and $\nudot$, several of which are even smaller in fractional size than those observed in \psr{0540}.  The Crab has, however, undergone several large glitch events, which have not yet been seen in \psr{0540}.
While this difference may be attributable to a variety of NS interior processes, it should be noted that the Crab pulsar has been observed with nearly four times as long a time baseline, and with much denser and more regular cadence, than has \psr{0540}.
It is also therefore possible that \psr{0540} has experienced unobserved large glitches, and/or may do so in the future.
\change{It should be noted that a $36\%$ increase has recently been observed in the spin down rate of \psr{0540}.  This is thought to be due to a state change perhaps caused by magnetospheric processes, rather than a glitch, since no accompanying rotation-rate change was seen \citep{mgh+15}.}

\begin{figure}
    \includegraphics[width=0.5\textwidth]{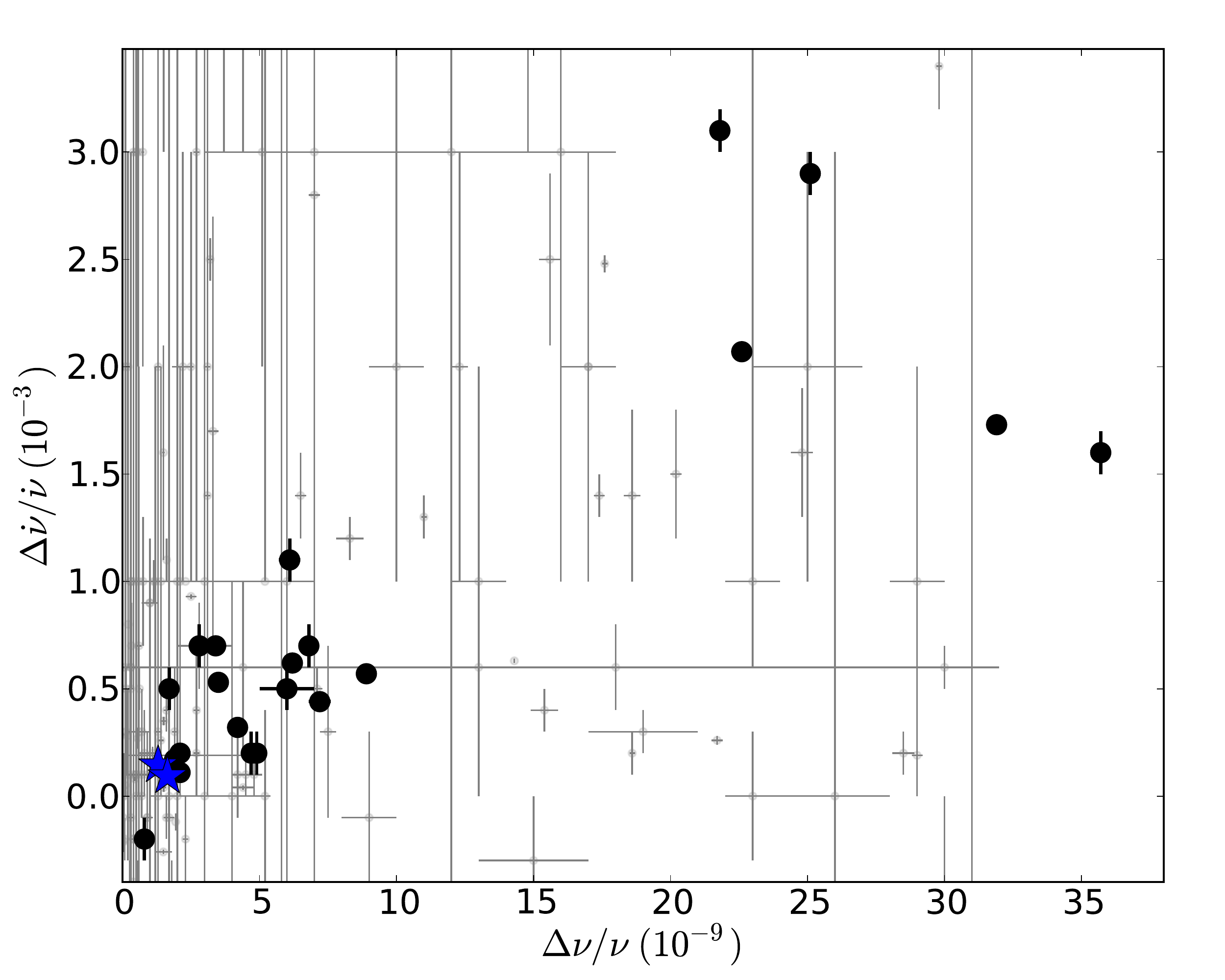}
    \caption{Fractional frequency and frequency derivative glitch sizes found for PSR~B0540$-$69 in this work (blue stars) compared with those found for the Crab pulsar \citep[black filled circles][]{ljg+15}, as well as all glitches (gray) reported in the glitch catalog published by \citet{elsk11} (and maintained online at \texttt{http://www.jb.man.ac.uk/pulsar/glitches.html}) that fall within the above plot limits.\label{fig:glitch_compare}}
\end{figure}

While the Crab has many more observed glitches than \psr{0540}, with 24 observed over 45 years \citep{elsk11,ljg+15}, both pulsars appear to follow the relatively low observed glitch rate of the very youngest pulsars \citep[e.g.,][]{sl96,elsk11}.  Along with the Crab and \psr{0540}, this group also includes PSR~J1119$-$6127 \citep[$\tau_c\sim1700$\,years;][]{ckl+00,wje11,ymh+13}, which has shown only 3 glitches in 16\,years of observations, and the similarly young pulsar, PSR~B1509$-$58 \citep[$\tau_c\sim1700$\,years;][]{kms+94,lkgm05,lk11}, which has not been observed to undergo a single glitch in observations spanning 28\,years.  The \change{low} glitch activity in \psr{0540} may therefore not be all that surprising, and may be related to its high internal temperature \citep{ml90}, compared to NSs with roughly an order of magnitude higher characteristic age, such as the Vela pulsar, which has had \change{significantly larger glitches}.   However, the relatively high \change{glitch activity} of magnetars, which appear to be the very hottest of all NSs, is not consistent with this hypothesis \citep{dkg08}.  Finally, it may also be possible that \psr{0540} underwent other low-level glitch events that, due to high-amplitude timing noise, were impossible to detect in earlier timing efforts \citep[see][for a thorough discussion of small glitches and timing noise]{eas+14}.  However, as seen in Figure~\ref{fig:glitch_compare}, the glitches we observe for \psr{0540} have fractional sizes that are consistent with the glitch size distribution of the Crab pulsar.

We can perform a rough comparison of glitch size and frequency between the Crab pulsar and \psr{0540} through the so-called glitch activity parameter $A_g = \sum({\Delta\nu/\nu})/\Delta{t}$, which quantifies the cumulative fractional change $\Delta\nu/\nu$ in spin frequency over the observing time span $\Delta{t}$ \citep{ml90}.  In the case of the Crab pulsar and \psr{0540}, we find $A_{g, \mathrm{Crab}} \sim 1\times10^{-8}\,$yr$^{-1}$ and $A_{g, \mathrm{0540}} \sim 0.02\times10^{-8}\,$yr$^{-1}$, based on the reported values from \citet{ljg+15} and this work, respectively.  It is fairly clear---apart from the unlikely possibility of an unobserved period of frequent and/or heavy glitch activity for \psr{0540} outside the dates observed in this work---that the Crab pulsar glitches are more frequent, and larger in magnitude, than those of \psr{0540}, and that this is one property in which these two objects differ significantly.

As can be seen in Figure~\ref{fig:part_coherent_summary}, $\nudot$ measurements in our partially coherent analysis show \change{persistent changes} after both glitches.
\change{This is also a feature of Crab pulsar glitches \citep[][]{girp77,dp83,ljg+15}, which also show persistent increases in spin-down, after a short ($\sim 100$-d) and incomplete recovery that terminates before reaching pre-glitch $\nudot$ values.  Recoveries are not seen in our \psr{0540} data, probably due to a much lower density data set, compared to the impressive observing cadence obtained for the Crab over 45 years \citep{ljg+15}.}
The persistent steps in $\nudot$ have been explained by models in which the effective NS moment of inertia is decreased due to the trapping of vortices in the interior, which are only released when a glitch occurs \citep{accp94}. An alternative model invokes the post-glitch movement of crustal ``plates'' \citep{rud91}, to which the surface magnetic field is coupled, causes an increased misalignment of the magnetic dipole moment from the spin axis, thus increasing the external torque and producing the observed offsets in $\nudot$ \citep{leb92}.

Finally, as with all measured braking indices, both the Crab pulsar and \psr{0540} have $n$ that is significantly less than 3, indicating a departure from the standard picture of magnetic dipole braking.  Along with a contribution from magnetospheric wind torques, one possible explanation involves the secular change in pulse profile component separation observed in the Crab pulsar, indicating a movement of the magnetic dipole axis away from the pulsar rotation axis, affecting its spin-down evolution \citep{lgw+13,ljg+15}.  However, the paucity of detected X-ray photon events for \psr{0540} requires us to choose a much coarser pulse profile binning in order to maintain sufficient S/N for timing analysis and emission studies.  This makes it nearly impossible to discern the subtle changes in pulse morphology that have been observed in the radio Crab pulsar profiles.  Furthermore, the weak radio flux of \psr{0540} \citep{mml+93} makes it difficult to perform this type of analysis at those wavelengths, as has been done with the Crab pulsar.  However, future radio telescopes and instrumentation with high projected sensitivity, such as the Square Kilometer Array \cite[e.g.,][]{wxe+15} may provide data sets that allow for such a study.

\section{Conclusions}
We have extended the timing analysis of the very young pulsar \psr{0540} to include a time span of nearly 16 years of \rxte{} data. We have confirmed the epoch and properties of the first glitch, previously reported by \citet{lkg05}, and discovered a second glitch.  We have confirmed that the braking index of this pulsar is consistent in value throughout our data set, and that no significant radiative changes have observed to be associated with its glitch activity, a behavior that is in line with the majority of young pulsars.   Many of the observed properties of \psr{0540}, such as fractional glitch depth, are consistent with several of those seen in the Crab pulsar. However, we do find that quantities describing other aspects of its behavior, such as the glitch rate, size distribution, and activity parameter, can differ substantially from the Crab, while showing greater similarity to those of other young pulsars.  In any case, there may be a variety of processes that contribute to the spin behavior of \psr{0540} and other recently born NSs. Further monitoring of these sources and discovery of more young pulsars will be crucial toward gaining a better understanding of their evolution and observed properties in the context of the overall NS population.

\acknowledgements
This research made use of data obtained from the High Energy Astrophysics Science Archive Research Center Online Service, provided by the NASA-Goddard Space Flight Center.  R.~F.~A. acknowledges support from an NSERC Alexander Graham Bell Canada Graduate  Scholarship and a Walter C.~Sumner Memorial Fellowship.  V.~M.~K. receives support from an NSERC Discovery Grant and Accelerator Supplement, Centre de Recherche en Astrophysique du Qu\'{e}bec, an R.~Howard Webster Foundation Fellowship from the Canadian Institute for Advanced Study, the Canada Research Chairs Program and the Lorne Trottier Chair in Astrophysics and Cosmology.  The authors would also like to thank M.~Livingstone for useful discussions and advice in the early stages of this analysis, C.~Espinoza for helpful comments, and the referee, whose comments have certainly improved the quality and clarity of this article.



\end{document}